\documentclass[10pt]{article}
\usepackage{jeosman}
\usepackage{amsmath,amssymb}
\usepackage{ae}

\usepackage{fancyhdr}
\begin{document}

%%%%%%%%%%%%%%%%%% title page information %%%%%%%%%%%%%%%%%%

\title{Active optics: deformable mirrors with a minimum number of actuators}
\vspace{0.4cm}
\author{M. Laslandes, E. Hugot, M. Ferrari}
\vspace{0.3cm}
\address{Aix Marseille Universit\'e, CNRS, LAM (Laboratoire d'Astrophysique de Marseille) \\ UMR 7326, 13388, Marseille, France}
\vspace{0.3cm}
\email{marie.laslandes@oamp.fr}

\vspace{1cm}
\textbf{Abstract}\\
We present two concepts of deformable mirror to compensate for first order optical aberrations.
Deformation systems are designed using both elasticity theory and Finite Element Analysis in order to minimize the number of actuators.
Starting from instrument specifications, we explain the methodology to design dedicated deformable mirrors.
The work presented here leads to correcting devices optimized for specific functions.
The Variable Off-Axis paraboLA concept is a 3-actuators, 3-modes system able to generate independently Focus, Astigmatism and Coma. 
The Correcting Optimized Mirror with a Single Actuator is a 1-actuator system able to generate a given combination of optical aberrations.\\
DOI: http://dx.doi.org/10.2971/jeos.2012.12036

\keywordline{active optics, deformable mirror, telescope, optical aberration, wave-front error correction, off-axis parabola}

%%%%%%%%%%%%%%%%%%%%%%%%%%  body  %%%%%%%%%%%%%%%%%%%%%%%%%%
\section{Active optics to control the wave-front}
\label{sec:intro}

\subsection{Active optics in astronomy}
\label{sec:intro_ao_astro}
Using deformable mirrors, active optics allows a wave-front control at nanometric precisions,
ensuring optimal performance for the optical instrument \cite{2000eaa..bookE4780.}.
For about twenty years, Earth-based telescopes have benefited from active optics systems, in three main domains.\\
The first application of active optics is the maintaining of large mirrors optimal shape with actuators located under their optical surfaces.
On Earth, the 8m-class telescopes have active primary mirrors, compensating for gravity effects and thermo-elastic deformations.
Developed and proved by Wilson on the ESO New Technology Telescope \cite{1987JMO....34..485W},
these systems are now widely used on 8-10m-class telescopes, such as Gemini North and South, Keck, Gran Telescopio Canarias (GTC) or Very Large Telescope (VLT).
For instance, the VLT 8.2 m primary mirror is maintained by 150 push/pull actuators  
\cite{1994SPIE.2199..271K}.\\
The second application consists in the use of dynamic optical components:
active mirrors are used in variable optical designs, to compensate for aberrations induced by moving elements.
Variable Curvature Mirrors (VCM), developed by Ferrari \cite{1998A&AS..128..221F}, provide this type of correction for the VLT
Interferometric mode.
The beams from the different telescopes are recombined through moving delay lines.
An efficient pupil stabilization is achieved with the application of a pressure under the VCMs' optical surfaces.\\
The third application is the generation of high optical quality aspherical mirror, using stress polishing.
Proposed in the 1930's by Schmidt \cite{Schmidt1932} for the polishing of the entrance correcting lens of his wide field telescope, 
this method has been improved by Lemaitre in 1974 \cite{1972ApOpt..11.1630L}.
It allows the achievement of an aspherical mirror without high spatial frequency errors,
by polishing a deformed optical substrate, under constraints, with a full-sized tool.
An interesting application of stress mirror polishing is the manufacturing of off-axis parabola for large segmented mirrors,
it has notably been used by Nelson for the manufacturing of the 36 segments of the Keck observatory primary mirrors \cite{1980ApOpt..19.2341N}.\\
As we can see through these three types of use, the key element of active optics is a deformable mirror,
designed and optimized to fit specific requirements.

\subsection{Evolution of deformable mirror and correction needs}
Active optics is complementary to adaptive optics.
On the one hand, the goal of adaptive optics is to correct high temporal and spatial frequency errors \cite{1998aoat.book.....H}.
On the other hand, the goal of active optics is to compensate for low order optical aberrations in the simplest and most efficient way.
The Wave-Front Errors (WFE) are classically decomposed on an orthonormal polynomials base, 
such as the Zernike polynomials, describing the optical aberrations \cite{1976JOSA...66..207N}.
In many cases, the correction need is limited to a few modes.
For Infra-Red and visible applications, the required precision of correction is on the order of tens nanometer ($\lambda/20$).\\
The problematic of maintaining mirrors' shapes begins to appear in space telescopes, large lightweight primary mirrors will be
sensitive to the environment variations and the induced thermo-elastic deformations will have to be compensated \cite{2011OptEn..50f3003C,1998SPIE.3356..758R}.
For a space use, the system simplicity and reliability are mandatory \cite{2010SPIE.7739E.105L}.\\
Stress polishing would also benefit from a simplification of warping systems.
Nowadays, the main application of this technique lies in the manufacturing of large mirrors' segments.
For instance, it is studied in the framework of the future European Extremely Large Telescope, for the mass production of the
thousand segments forming the primary mirror \cite{2011SPIE.8169E...2L}.
In such a case, it is important to have a simple deformation system, optimized for the required optical shape, improving the process efficiency.\\
Finally, active systems optimization finds a direct application for variable optical components.
As well as the thermo-elastic deformation can be predicted with Finite Element Analysis, 
the aberrations induced by an instrument reconfiguration can be easily anticipated in the early instrument design phases, using ray-tracing software.
Knowing the wave-front errors that would have to be compensated, a dedicated correcting system can be conceived.\\
In this paper, we present active systems optimized to the extreme: an optical mode is corrected with a single actuation point.
The interest of such a minimization lies in the set-up and monitoring easiness, but also in the limited weight and power consumption 
of the systems.\\
The design of deformation system is based on the elasticity theory, describing the mechanical behavior of plates under
given boundary conditions \cite{Timoshenko1959}.
A mirror can be deformed by many ways, such as the application of forces, displacements, bending moments or pressure \cite{Lemaitre2009}.
There are also many actuator technologies: mechanic, magnetic, electric or piezoelectric.\\
Some works present deformable mirrors with a limited number of actuators but none have pushed the limits to a single actuator.
For instance, Dainty has developed a 9 channels bimorph deformable mirror, to generate focus, astigmatism, coma and spherical
aberrations \cite{1998ApOpt..37.4663D},
and Freeman presents a 3-actuators deformable mirror able to generate focus and astigmatism \cite{1982ApOpt..21..580F}.
In these systems, the forces are directly applied on the optical surface,
inducing a print-through effect.
The resulting high spatial frequency errors deteriorate the correction performance.\\
In the work presented here, the number of actuator is minimized by deporting the forces far from the optical surface.
In this manner, the number of actuators is uncoupled from the mirror diameter and the generation of high spatial
frequency errors is avoided.
Moreover, it dissociates the actuator precision from the shape generation precision and
it makes the deformable mirror independent of the actuator technology.
By coupling elasticity equations and Finite Element Analysis, we detail hereafter innovative and simple concepts to warp mirrors.

\section{Adapting the influence functions to the correction requirements}
\label{sec:method}
The concepts developed for the minimization of the number of actuators are based on the elasticity
theory, and particularly on the deformation of a thin shell through the application of bending moments at its edges,
as described by Timoshenko \cite{Timoshenko1959}.
The input for the optimization is the correction requirement.
The goal is then to match the system's influence functions with the optical modes to be corrected, in order to have one actuator per mode.\\
The first concept, Variable Off-Axis paraboLA (VOALA), allows the generation of focus, astigmatism and coma with three actuators, 
each actuator driving a mode \cite{VOALA}.
The second concept, Correcting Optimized Mirror with a Single Actuator (COMSA), allows the generation of a given combination of aberrations
with only one actuator \cite{COMSA}.
\subsection{Principle: mirror curvature modification}
\label{sec:method_curvature}
\begin{figure}[!h]
 \centering
  \includegraphics[scale=0.5]{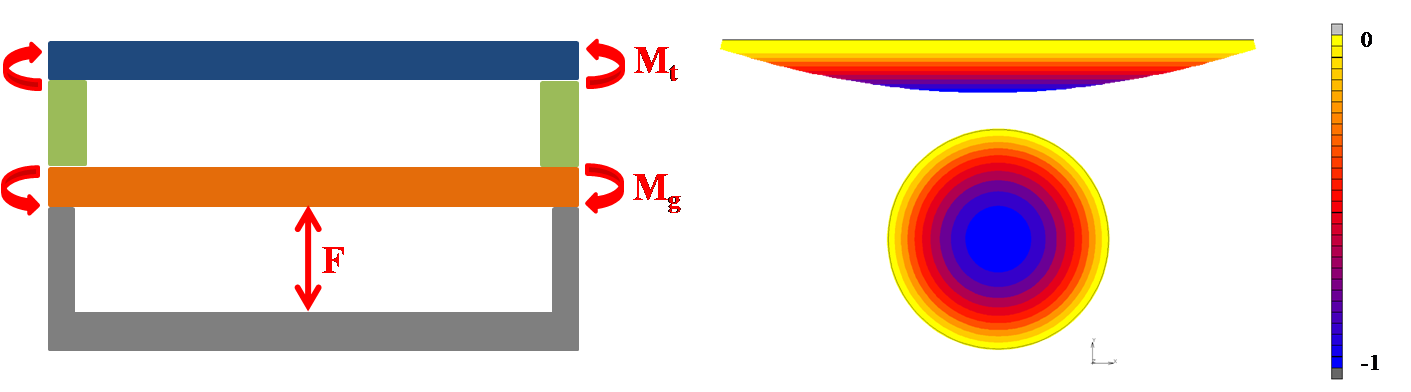}
  \caption{Left: Transverse section of a circular system generating Focus (blank in blue, intermediate plate in orange, ring in green and clamping system in grey).
The application of a force $F$ on the intermediate plate induces bending moments at the plate edges ($M_g=Fa$), which are transmitted to the mirror edges ($M_t=M_g$).
Right: Views of the deformation induced on the optical surface (from FEA).}
  \label{fig:gener_foc}
\end{figure}

Both concepts are based on the modification of an optical surface curvature through
the application of uniform bending moments at its edges.
These moments are generated on an intermediate plate with a central force and transmitted to the mirror via a flexible outer ring.
As it can be seen on Fig. \ref{fig:gener_foc}, such a system is constituted of five main parts:
three plates (mirror, intermediate plate and rigid reference plate) and two rings linking the plates together.
The central actuator, applying a force or a displacement on the system, is located between the intermediate and reference plates.\\
On a circular blank of semi-diameter $a$, the moments generated at the intermediate plate edges by the central force $F$ are constant: 
$M_g=Fa$.
These moments are transmitted at the edges of the mirror: $M_t=M_g$.
The induced mirror deformation $z_{mirror}$ is deduced from the elasticity equation describing this load-case
(application of bending moments $M_t$ at the edges of an axisymmetric clamped plate) \cite{Timoshenko1959}:
\begin{equation}
 D\frac{{\partial}^2z_{mirror}}{{\partial}r^2}=-M_t,
\end{equation}
with $D$ the plate rigidity, $\nu$ the Poisson ratio and the following boundary conditions:
\begin{equation}
z_{mirror}(r=a)=0 \;\;\; and \;\;\; \frac{{\partial}z_{mirror}}{{\partial}r}(r=0)=0.
\end{equation}
The resolution of this differential equation conducts to a focus mode:
\begin{equation}
z_{mirror}(r)=-\frac{M_t}{2D}(r^2-a^2).
\end {equation}
We will see thereafter how to adapt this principle to generate other aberrations.
\subsection{Design and analysis methods}
\label{sec:method_design}
Minimizing the number of actuators is only possible with a strong preliminary work to specify the correction requirements.
Starting with an instrument design and operating environment, the Wave-Front Error (WFE) can be predicted
and decomposed on optical modes.
An optical mode is defined as a given combination of optical aberrations (such as focus, astigmatism or coma).
The problematic is then to design an active mirror compensating for the expected WFE with a single actuator for each
required optical mode.
This is achieved by adapting the geometry of the system described in Fig. \ref{fig:gener_foc}.\\
The elasticity equations give the right bending moments to apply at the edges of the mirror in order to generate the
required optical surface deformation.
It defines the main system characteristics.
Then, Finite Element Analysis (FEA) is performed for optimization \cite{Smith2004}, considering both optical quality and mechanical system strength.
The optimization output parameters are the system dimensions, its materials and the actuators characteristics while the
optimization criteria are the optical surface shape $z_{mirror}$, compared to the required deformation $z_{required}$,
but also the maximum level of stress in the material ${\sigma}_{max}$, and the required force $F$.
A classical least square approach \cite{Bonnans2009} is used to converge to the deformation system design minimizing the quantity $\alpha$
defined in Eq. \ref{eq:min}:
\begin{equation}
 \alpha(geometry,material,actuator)={\lambda}_{res}{\parallel{z_{mirror}-z_{required}}\parallel}^2+{\lambda}_{\sigma}{\sigma}_{max}+{\lambda}_F{F},
\label{eq:min}
\end{equation}
where $z_{mirror}-z_{required}$ is the residual deformation and $\lambda$ is a weight given to each parameter.
The deformable mirror performance is characterized by its precision $p$, defined as:
\begin{equation}
 p={\parallel{z_{mirror}-z_{required}}\parallel}^2/{\parallel{z_{required}}\parallel}^2.
\label{eq:prec}
\end{equation}

\section{Off-Axis Parabola generation with a 3-actuators system}
\subsection{Application domain}
\label{sec:voala}
An Off-Axis Parabola is characterized by its pupil semi-diameter $a$, radius of curvature $k$, conic constant $C$ and off-axis distance $R$.
As described by Lubliner \& Nelson \cite{1980ApOpt..19.2332L}, the optical surface shape $z_{mirror}$, can be deduced from these parameters.
It is composed of a sphere plus terms corresponding to the first optical aberrations:
\begin{equation}
z_{mirror}(\rho,\theta,a,k,R,C)=\sum{{\alpha}_{ij}(a,k,R,C)Z_{ij}(\rho,\theta)}
\label{eq:oap}
\end{equation}
where ${\alpha}_{ij}$ are the optical modes amplitudes and $Z_{ij}$ the Zernike polynomials, described in Table \ref{table:aberration_oap}.
\begin{table}[h]
 \caption{Equations of Zernike coefficients and polynomials as a function of OAP characteristics (with the reduced parameter $\epsilon=R/k$).}
\label{table:aberration_oap}
\begin{center}       
\begin{tabular}{|l|l|l|} 
\hline
\rule[-1ex]{0pt}{3ex} \small{} & ${\alpha}_{ij}$ \small{(RMS) - first order approximation} & \small{$Z_{ij}$}\\
\hline
\rule[-1ex]{0pt}{3ex} \small{Focus} & ${\alpha}_{20}=\frac{a^2}{2\sqrt{3}k}\frac{2-C{\epsilon}^2}{4(1-C{\epsilon}^2)^{3/2}}$ & \small{$Z_{20}=\sqrt{3}(2\rho^2-1)$}\\
\hline
\rule[-1ex]{0pt}{3ex} \small{Astm3} & ${\alpha}_{22}=\frac{a^2}{\sqrt{6}k}\frac{C{\epsilon}^2}{4(1-C{\epsilon}^2)^{3/2}}$ & \small{$Z_{22}=\sqrt{6}\rho^2cos(2\theta)$} \\
\hline
\rule[-1ex]{0pt}{3ex} \small{Coma3} & ${\alpha}_{31}=\frac{a^3}{3\sqrt{8}k^2}\frac{C{\epsilon}[1-(C+1){\epsilon}^2]^{1/2}(4-C{\epsilon}^2)}{8(1-C{\epsilon}^2)^{3}}$ & \small{$Z_{31}=\sqrt{8}(3\rho^2-2)\rho{cos(\theta)}$} \\ 
\hline
\rule[-1ex]{0pt}{3ex} \small{Tref5} & ${\alpha}_{33}=\frac{a^3}{\sqrt{8}k^2}\frac{C^2{\epsilon}^3[1-(C+1){\epsilon}^2]^{1/2}}{8(1-C{\epsilon}^2)^{3}}$ & \small{$Z_{33}=\sqrt{8}\rho^3cos(3\theta)$} \\
\hline
\rule[-1ex]{0pt}{3ex} \small{Sphe3} & ${\alpha}_{40}=\frac{a^4}{6\sqrt{5}k^3}\frac{8(C+1)-24K{\epsilon}^2+3C^2{\epsilon}^4(1-3C)-C^3{\epsilon}^6(2-C)}{64(1-C{\epsilon}^2)^{9/2}}$ & \small{$Z_{40}=\sqrt{5}(6\rho^4-6\rho^2+1)$}\\
\hline
\rule[-1ex]{0pt}{3ex} \small{Astm5} & ${\alpha}_{42}=\frac{a^4}{4\sqrt{10}k^3}\frac{-C{\epsilon}^2(1+5C-C{\epsilon}^2(6+5C)}{16(1-C{\epsilon}^2)^{7/2}}$ & \small{$Z_{42}=\sqrt{10}(4\rho^2-3)\rho^2cos(2\theta)$} \\
\hline
\end{tabular}
\end{center}
\end{table}
\\
Starting with the equations giving the amplitude of each Zernike polynomial as a function of the OAP characteristics, 
we can define a domain where the first three aberrations are predominant compared to the others.
We consider that Trefoil5, Spherical3 and Astigmatism5 are negligible when their amplitudes are lower than a given threshold
$a_{th}$ which depends on the considered application.
It gives conditions on quadruplets (a,k,C,R):
\begin{equation}
\begin{cases}
 {\alpha}_{33}<a_{th} => a<\Big{[}a_{th}\sqrt{8}k^2\frac{8(1-C{\epsilon}^2)^{3}}{C^2{\epsilon}^3[1-(C+1){\epsilon}^2]^{1/2}}\Big{]}^{1/3} \\
 {\alpha}_{40}<a_{th} => a<\Big{[}a_{th}6\sqrt{5}k^3\frac{64(1-C{\epsilon}^2)^{9/2}}{8(C+1)-24C{\epsilon}^2+3C^2{\epsilon}^4(1-3C)-C^3{\epsilon}^6(2-C)}\Big{]}^{1/4} \\
 {\alpha}_{42}<a_{th} => a<\Big{[}a_{th}4\sqrt{10}k^3\frac{16(1-C{\epsilon}^2)^{7/2}}{-C{\epsilon}^2(1+5C-C{\epsilon}^2(6+5C)}\Big{]}^{1/4}.
\end{cases}
\end{equation}
The Variable Off-Axis paraboLA concept is a 3-actuators deformation device designed to generate any combination of Focus,
Astigmatism3 and Coma3, in the limit of actuators stroke and system's mechanical strength.
Thus, for a given maximal amplitude of residual aberrations, the set of OAPs achievable with the VOALA system can be defined.
\subsection{Focus and coma generation}
\label{sec:voala_focus_coma}
\begin{figure}[!h]
 \centering
  \includegraphics[scale=0.39]{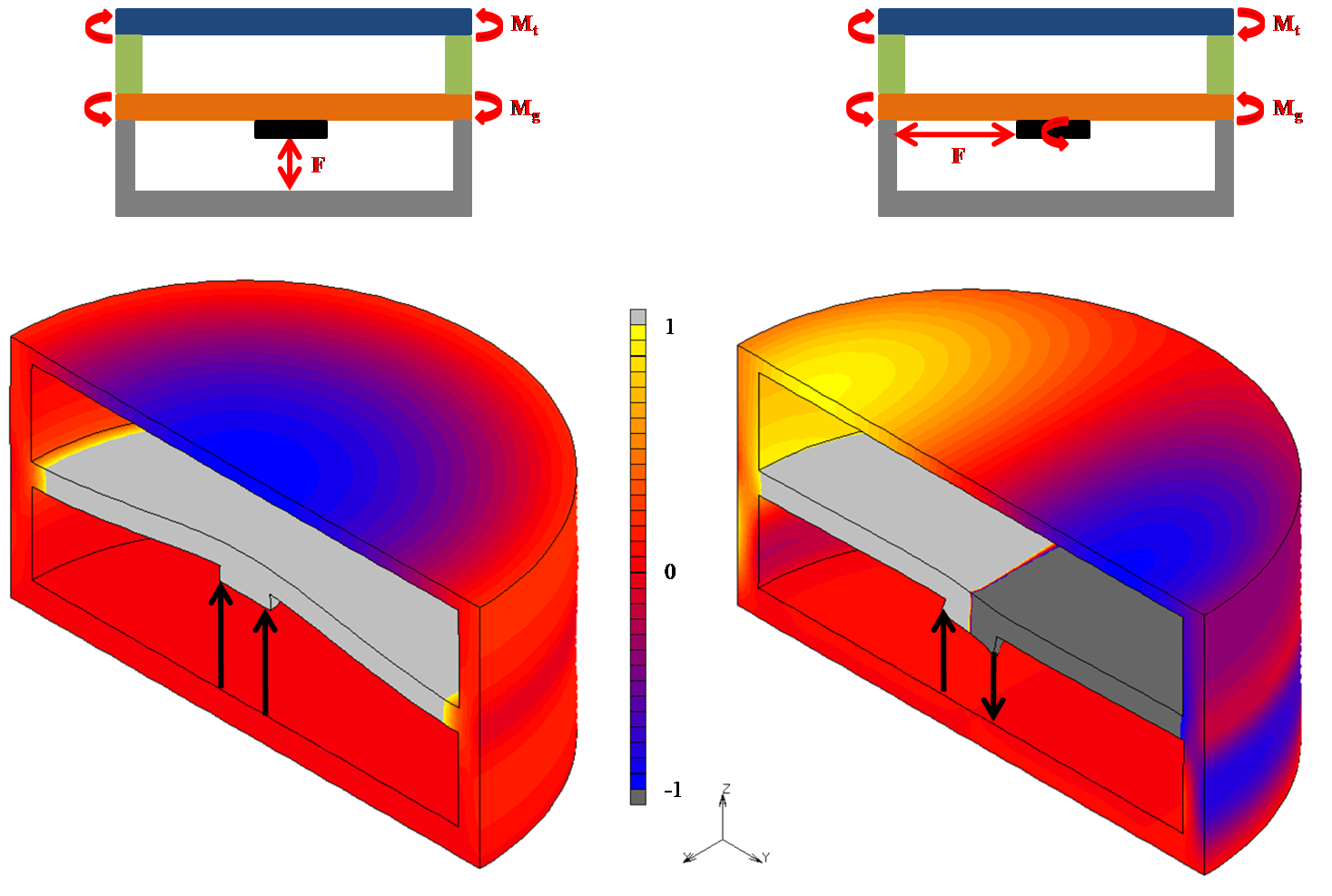}
  \caption{Top: Principle of generation of Focus (left) and Coma (right) with the VOALA concept.
Bottom: FEA model of the 2-actuators system presenting the deformations obtained with the load cases corresponding to Focus and Coma generations.}
  \label{fig:voala_foc_coma}
\end{figure}

As explained in Section \ref{sec:method_curvature}, a circular plate can be deformed in a focus mode by applying 
constant uniform bending moment at its edges.
These moments are generated on an intermediate plate with a central force.
On the same principle, we search for the moment distribution to apply at the mirror edges to generate a coma mode.
The link between bending moments and plate deformation is given by the elasticity theory \cite{Timoshenko1959}:
\begin{equation}
 \label{eq:M_vs_def}
M_t(r,\theta)=-D[\frac{\partial^2{z(r,\theta)}}{\partial{r^2}}+\nu(\frac{1}{r}\frac{\partial{z(r,\theta)}}{\partial{r}}+\frac{1}{r^2}\frac{\partial^2{z(r,\theta)}}{\partial{\theta^2}})],
\end{equation}
with $D$ the plate rigidity and $z(r,\theta)$ the required plate deformation in cylindrical coordinates:
\begin{equation}
 \label{eq:def_coma}
z(r,\theta)=\Big{[}\Big{(}\frac{r}{a}\Big{)}^2-1\Big{]}\frac{r}{a}cos(\theta),
\end{equation}
Combining Eq. \ref{eq:M_vs_def} and \ref{eq:def_coma}, we obtained the expression of bending moments at the mirror edges:
\begin{equation}
 \label{eq:moment_coma}
M(r=a,\theta)=\frac{-2D}{a^2}(3+\nu)cos(\theta).
\end{equation}
We deduce that the application of an azimuthal moment distribution at the mirror edges induces a coma mode on the optical surface.\\
As described by Timoshenko, such a moment distribution is achieved through the application of a
central mechanical moment on the intermediate plate.
To generate the central moment, a central pad is added on the intermediate plate and a transverse force is applied on this pad.
Figure \ref{fig:voala_foc_coma} presents the required load cases on the system to generate focus and coma.
Combining the actions of two actuators located on a pad diameter, both modes can be generated.
If the two forces are equal, it corresponds to a central force application and leads to a focus.
If the two forces are opposite, it corresponds to a central moment application and leads to a coma.
Other forces configurations correspond to combinations of focus and coma.
\subsection{Astigmatism generation}
\label{sec:voala_astm}
\begin{figure}[!h]
 \centering
  \includegraphics[scale=0.45]{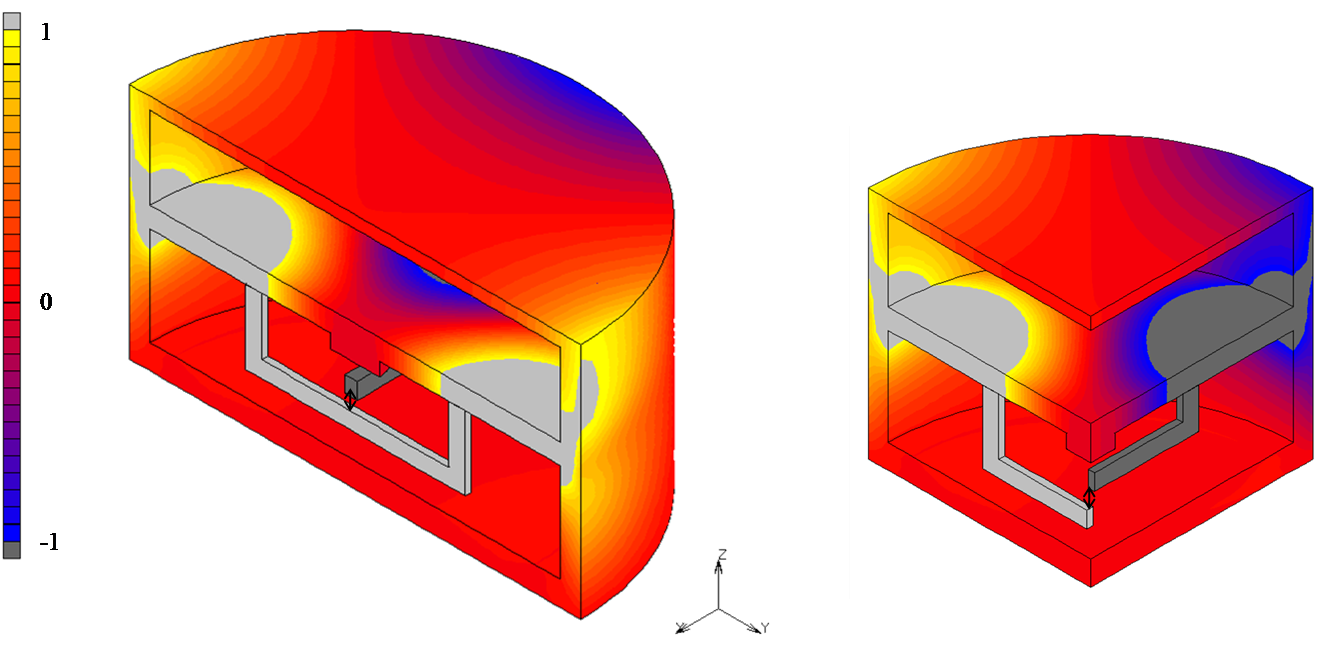}
  \caption{FEA model presenting the principle of Astigmatism generation with VOALA.}
  \label{fig:voala_astm}
\end{figure}

The principle of astigmatism generation with one actuator, developed by Hugot \cite{2008ApOpt..47.1401H}, can be added to the previous system.
An astigmatism figure can be generated by applying two pairs of opposite forces on two orthogonal diameters of the intermediate plate.
On the finite element model, we study the evolution of the deformation in function of the forces diameter.
The optimum diameter is the one minimizing the residual deformation.
The four forces can be generated from a single point, located between two rigid orthogonal beams.
Each beam is linked to two points of the diameter.
Applying a central force pushing aside the two beams, the required forces are transmitted on the four points.\\
Figure \ref{fig:voala_astm} presents the FEA model for this astigmatism generation.
We can note that the two-beams system can conveniently be installed on either side of the intermediate plate,
depending on the available space.
\subsection{Modes combination and alternative design}
\label{sec:voala_combi}
\begin{figure}[!h]
 \centering
  \includegraphics[scale=0.5]{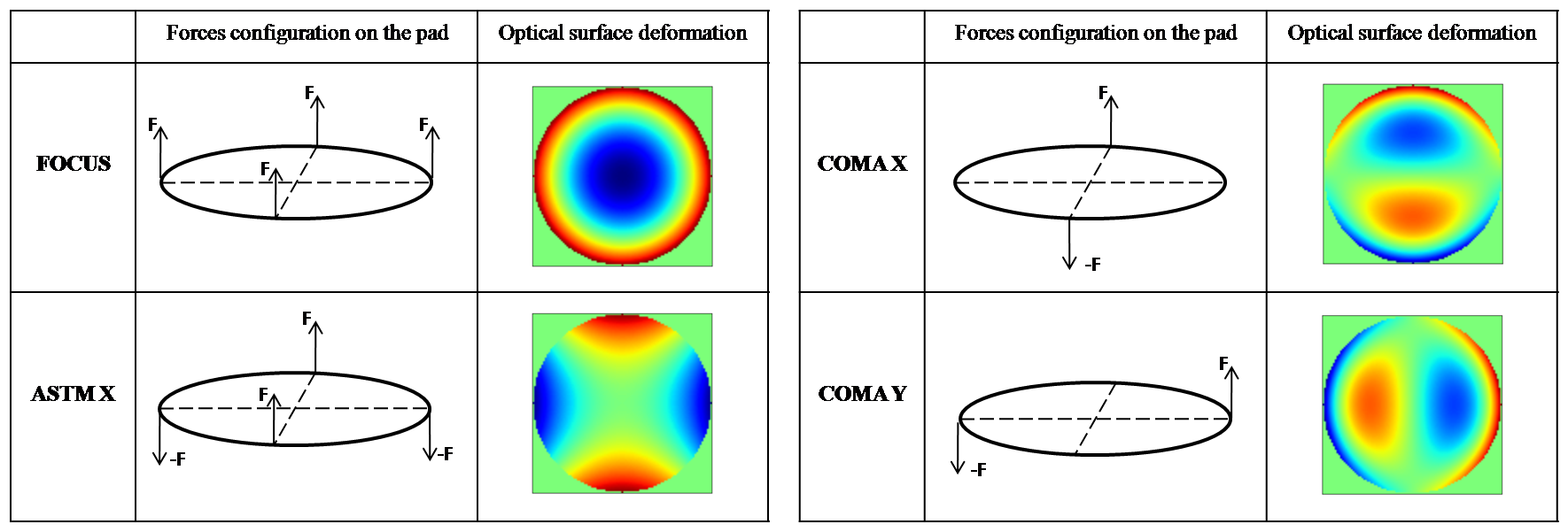}
  \caption{Variant of the VOALA design: generation of 4 Zernike polynomials with 4 actuators on the pad.
A fifth actuator allowing a system rotation can be added in order to generate the second astigmatism mode.}
  \label{fig:voala_4act}
\end{figure}

The 3-actuators system described above is able to compensate for combinations of focus, astigmatism and coma.
One actuator directly drives the astigmatism generation while focus and coma are generated
with a combination of the two other actuators.
But with this system, coma and astigmatism are oriented. 
Astigmatisms in both x and y directions could be generated with an additional beams system, turned of $\pi/4$ in comparison to the first one.
It is also conceivable to integrate the deformable mirror on a rotating platform, driven by one actuator,
the system rotation will then allow the generation of both astigmatisms.\\
An interesting alternative appears with the generation of the two comas:
with 4 actuators, located on the pad, on two orthogonal diameters, focus, coma x, coma y and astigmatism x (or y) can be created.
Figure \ref{fig:voala_4act} presents the load cases for the generation of each mode.
As explained just above, a fifth actuator rotating the system would provide the last astigmatism. 
This improvement leads to a 5 actuators - 5 modes deformable mirror.
\newpage

\section{Single actuator deformable mirror}
\label{sec:comsa}                                        
In some instruments, the wave-front to be corrected and its evolution can be predicted using ray-tracing models.
If the correction need is a combination of optical aberrations, evolving linearly with time, 
the Wave-Front Error (WFE) can be defined as a composite optical mode:
\begin{equation}
 WFE(t)=A(t)\sum{{\alpha}_{ij}Z_{ij}},
\end{equation}
with $Z_{ij}$ given Zernike polynomials, ${\alpha}_{ij}$ their initial amplitudes and $A(t)$ a coefficient giving their evolution with time.\\
For instance, correcting mirrors can be used to compensate for Optical Path Difference (OPD) in off-axis interferometers,
where differential aberrations in the instrument arms will be focus, astigmatism and eventually coma and tilt,
evolving linearly with the interferometer arms length.
It is then possible to adapt the single actuator concept presented in Section \ref{sec:method_curvature} to generate the required deformation
(see Fig. \ref{fig:gener_foc}).
The design method consists in modifying the system geometry to match the actuator influence function with the correction need.
The three parameters to be optimized are the system contour, the intermediate plate thickness distribution and the actuator location. 
\subsection{Contour adaptation}
\label{sec:comsa_contour}
\begin{figure}[!h]
 \centering
  \includegraphics[scale=0.35]{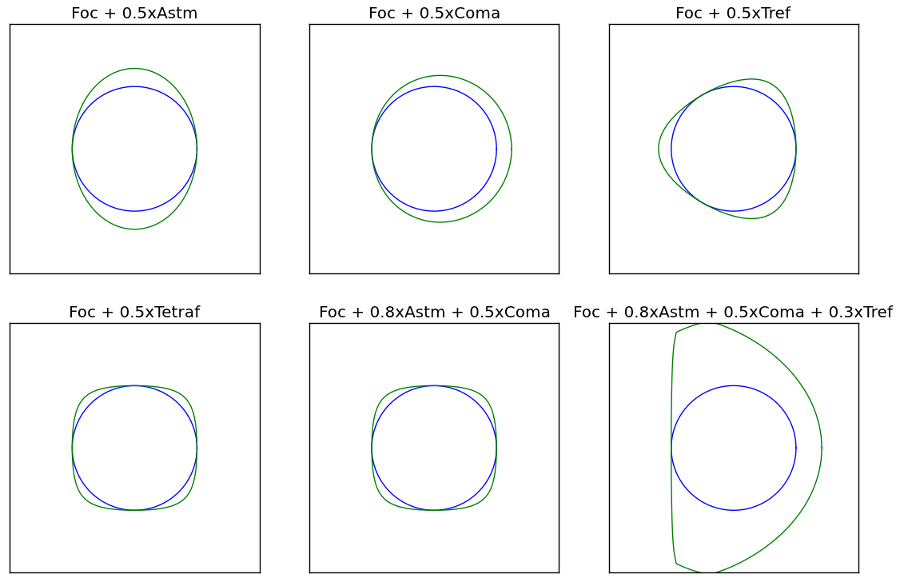}
  \caption{Contours for different optical modes combinations (circular pupil contour in blue).}
  \label{fig:contour}
\end{figure}

An angular modulation of the radius of curvature can be achieved by modifying the system contour according
to the combination of modes to be generated.\\
The contour $\rho_c$ is defined as a function of the required deformation on the circular pupil: 
$z(\rho,\theta)=\sum{{\alpha}_{ij}Z_{ij}(\rho,\theta)}$.
The system's boundary condition defines the contour: the mirror has clamped edges.
So, the required deformation is extended until it crosses the $z=0$ plane and the system contour corresponds to the intersection 
between this plane and the deformation surface.
This contour is expressed as a function of the angular coordinate $\theta$, and the modes amplitudes ${\alpha}_{ij}$,
as described in Eq. \ref{eq:contour}:
\begin{equation}
\label{eq:contour}
z(\rho_c,\theta)=\sum{{\alpha}_{ij}Z_{ij}(\rho_c,\theta)}=0   =>   \rho_c=f(\theta,{\alpha}_{ij}).
\end{equation}
Figure \ref{fig:contour} presents examples of contour computed to generate given combinations of aberrations.
\subsection{Thickness distribution}
\label{sec:comsa_thickness}
The moment distribution at intermediate plate edges $M_g$, is generated with a central force on this plate.
So, the system contour defines the bending moment modulation:
\begin{equation}
 \label{eq:Mint}
M_g(\theta)=F\rho_c(\theta)=F\frac{r_c(\theta)}{a},
\end{equation}
where $F$ is the applied force, $r_c(\theta)$ is the distance from the center to the edge for a given orientation and $a$ is the optical pupil radius.\\
On the other hand, as seen in Section \ref{sec:voala_focus_coma}, the bending moment modulation $M_t$ 
to apply at mirror edges to produce the required deformation, $z(r,\theta)$ is given by the elasticity theory \cite{Timoshenko1959}:
\begin{equation}
 \label{eq:Mdef}
M_t(\theta)=-\frac{Et^3}{12(1-\nu^2)}[\frac{\partial^2{z(r_c,\theta)}}{\partial{r_c(\theta)}^2}+\nu(\frac{1}{r_c(\theta)}\frac{\partial{z(r_c,\theta)}}{\partial{r_c(\theta)}}+\frac{1}{r_c(\theta)^2}\frac{\partial^2{z(r_c,\theta)}}{\partial{\theta^2}})],
\end{equation}
where $t$ is the plate thickness, $E$ its Young modulus and $\nu$ its Poisson ratio.\\
The generated moments $M_g$ are transmitted to the mirror edges ($M_t$) and induce the optical surface deformation.
Solving $M_g(\theta)=M_t(\theta)$ gives the angular thickness distribution of the intermediate plate, $t_c(\theta)$,
generating the required bending moments with the system contour: 
\begin{equation}
\label{eq:thickness}
t_c(\theta)=[\frac{12(1-\nu^2)F}{E}\frac{r_c(\theta)}{a}(\frac{\partial^2{z(r_c,\theta)}}{\partial{r_c(\theta)}^2}+\nu(\frac{1}{r_c(\theta)}\frac{\partial{z(r_c,\theta)}}{\partial{r_c(\theta)}}+\frac{1}{r_c(\theta)^2}\frac{\partial^2{z(r_c,\theta)}}{\partial{\theta^2}}))^{-1}]^{1/3}. 
\end{equation}
\subsection{Actuator location}
\label{sec:comsa_act}
The last system parameter is the force location.
We have seen in Eq. \ref{eq:Mint} that the transmitted bending moments depend on the distance between the force location and the edges.
Considering a decentering of $(x_d,y_d)$, the new distance $r'_c(\theta)$ induces a new bending moment modulation and Eq. \ref{eq:Mint} becomes:
\begin{equation}
 \label{eq:Mint_decentre}
M_{g}(\theta)=F\frac{r'_c(\theta)}{a}=\frac{F}{a}\sqrt{r_c(\theta)^2+x_d^2+y_d^2-2r_c(\theta)(x_dcos(\theta)+y_dsin(\theta))}.
\end{equation}
The induced modulations in $cos(\theta)$ and $sin(\theta)$ correspond to a generation of tilt and coma, 
their amplitudes depending on the shifting distance.
So, these two modes can be generated either by defining a specific contour or by decentering the actuator.
In order not to damage the other modes quality, the thickness distribution can be recalculated
equalizing Eq. \ref{eq:Mdef} and \ref{eq:Mint_decentre}.

\section{Systems performance analysis with a study case}
\label{sec:perfo}
In order to use the deformable mirrors in optical systems for visible or infra-red observations, 
the systems' residual wave-front error is specified around $\lambda/20$.
Thus, our goal is to design deformation systems accurate to a few tens of nanometers.\\
As a study case, we consider the generation of a 100 mm diameter Off-Axis Parabola,
with a focal ratio of 1, and a $45^\circ$ off-axis angle.
Using the formulas describing an OAP \cite{1980ApOpt..19.2332L}, such an optical shape can be decomposed as a sphere of radius 200 mm
plus terms of focus (${\alpha}_{20}=1.73$ {$\mu$}m rms), astigmatism3 (${\alpha}_{22}=2.03$ {$\mu$}m rms) and coma3 (${\alpha}_{31}=0.35$ {$\mu$}m rms).\\
Both systems are designed as described in the previous sections and we present hereafter their performance, computed with Finite Element Analysis.
For this example, the systems are simulated as monolithic pieces.
The material chosen is aluminum, for its flexibility (Young's modulus $E=75$ GPa, Poisson ratio $\nu=0.33$).
Both finite element models have 115230 hexaedral elements and 139350 nodes,
with a sampling of 7600 nodes on the optical surface, allowing a good characterization of the optical quality.
The back face of the reference plate is fixed and the forces are applied on the intermediate plate to simulate the actuators.
For a prototyping it is planned to assemble 3 parts together:
the central part will consist in the intermediate plate with the 2 rings,
the mirror will be glued on the top ring and the reference plate will be glued on the bottom ring (see Figure \ref{fig:gener_foc}).
From the Finite Element Analysis, the systems' performance are characterized by computing
the optical precision, defined as the difference between required and generated shapes.

\subsection{VOALA performance}
\label{sec:perfo_voala}
\begin{figure}[!h]
 \centering
  \includegraphics[scale=0.25]{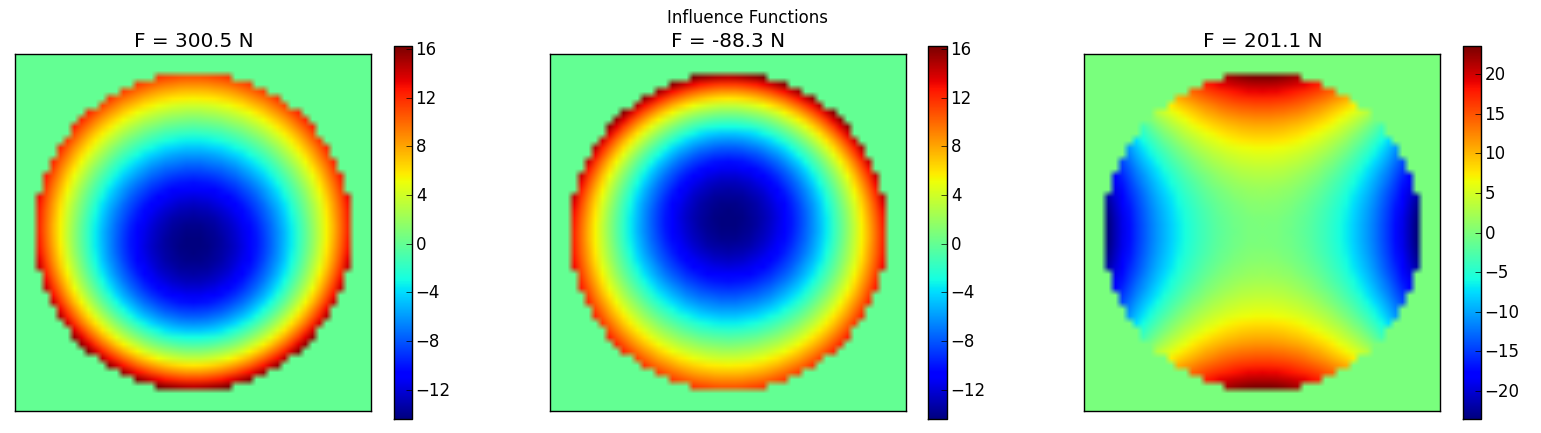}
  \caption{VOALA system influence functions (1st actuator on the central pad - 2nd actuator on the central pad - 3rd actuator between the two beams)
and their projection coefficients to generate the required OAP (deformation maps unit = nm) (FEA results).}
  \label{fig:FI_VOALA}
\end{figure}

\begin{figure}[!h]
 \centering
  \includegraphics[scale=0.25]{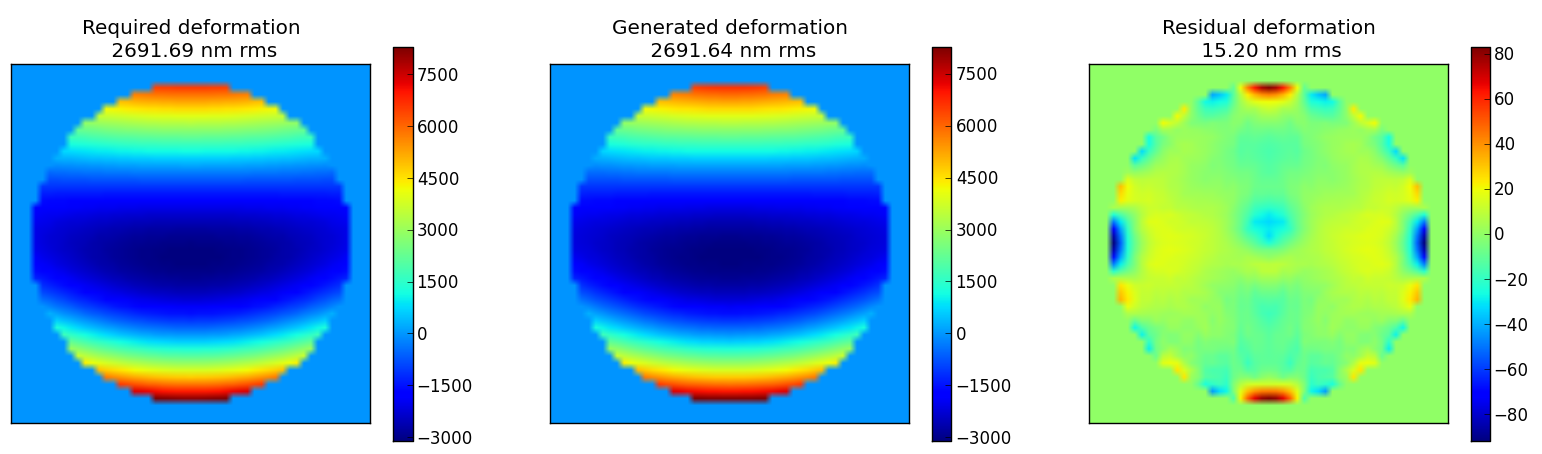}
  \caption{VOALA system performance for the OAP generation: Required deformation - Deformation generated on the optical surface with the 3 actuators - Residual deformation (Unit = nm) (FEA results).}
  \label{fig:perfo_voala}
\end{figure}

The required mirror shape is decomposed on the three system's influence functions, recovered on the Finite Element model:
it gives the values of the forces to be applied by each actuator (see Fig. \ref{fig:FI_VOALA}).
As we can see in Fig. \ref{fig:perfo_voala}, the mirror shape is generated with a high precision:
the residual deformation is 15.2 nm rms and is mainly composed of astigmatism harmonics.\\
Such a system is interesting to generate different OAPs, ie different combinations of focus, astigmatism and coma.
To determine an optimal range of use, we characterize the generation of each mode separately (see Table \ref{table:perf_voala}).
With a 0.2 \% precision the focus generation is highly efficient and it does not require a lot of force, so it is 
not a critical mode.
Astigmatism is generated with 0.7 \% precision, the required force is almost twice higher than for the focus but it is split
among 4 points on the system so this mode will not induce too much mechanical stress either.
As the required amplitude of coma was lower of one order of magnitude compared to the other modes,
its weight in the system optimization was less important.
It leads to a generation of coma slightly less efficient: 1.1 \% of residual deformation with higher forces required on the actuators.
\begin{table}[!h]
 \caption{System performance for the generation of 1 $\mu$m rms of each mode with the VOALA system (FEA results).}
\label{table:perf_voala}
\begin{center}       
\begin{tabular}{|c|cccc|} 
\hline
\rule[-1ex]{0pt}{3ex}   & Act1 (pad) & Act2 (pad) & Act3 (beams) & Residues\\
\rule[-1ex]{0pt}{3ex}   & (N) & (N) & (N) & (nm rms)\\
\hline
\rule[-1ex]{0pt}{3ex}   FOCUS & 61.3 & 61.3 & 0 & 1.92 \\
\rule[-1ex]{0pt}{3ex}   COMA & 555.4 & -555.4 & 0 & 10.94 \\
\rule[-1ex]{0pt}{3ex}   ASTM & 0 & 0 & 99.9 & 7.01 \\
\hline
\end{tabular}
\end{center}
\end{table}
\newpage

\subsection{COMSA performance}
\label{sec:perfo_comsa}
\begin{figure}[!h]
 \centering
  \includegraphics[scale=0.3]{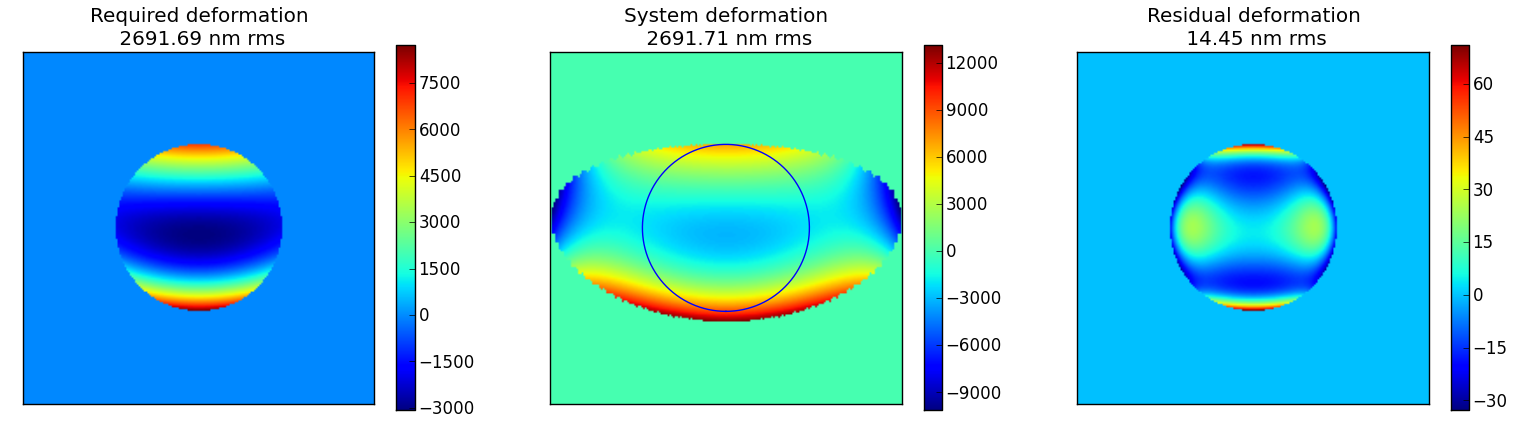}
  \caption{COMSA system performance: Required deformation (deduced from the OAP characteristics) - Optical surface deformation (generated by the actuator force) - Residual deformation on the pupil (Unit = nm) (FEA results).}
  \label{fig:perfo_OAP}
\end{figure}

\begin{figure}[!h]
 \centering
  \includegraphics[scale=0.3]{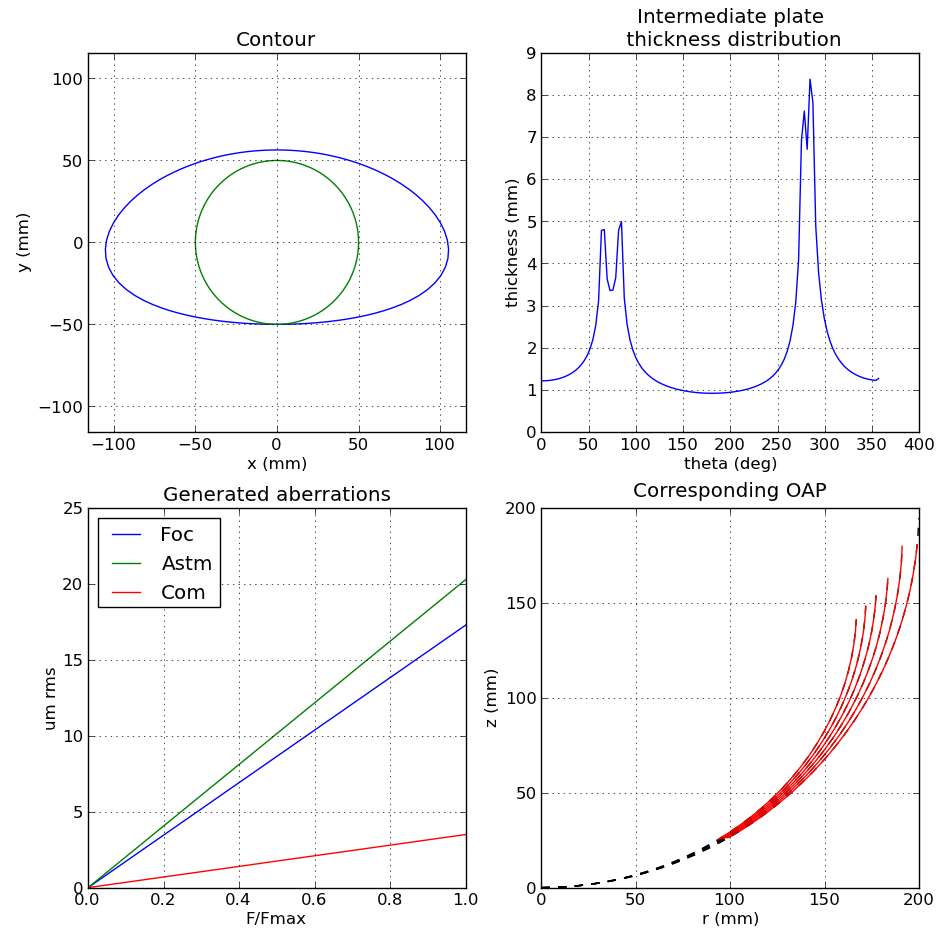}
  \caption{Analytical solution for OAP generation with COMSA: System contour and thickness distribution defined by the required OAP - Aberrations achievable with such a system - Corresponding OAP.}
  \label{fig:OAP_comsa}
\end{figure}

Starting with the given combination of optical aberrations, we apply the methodology explained in Section \ref{sec:comsa}
to design the COMSA system.
The computed contour and thickness distribution are presented in Fig. \ref{fig:OAP_comsa}.
This Figure also shows the combinations of aberrations achievable with such a system and the corresponding OAP:
the optical shape evolves linearly with the value of the applied force:
\begin{equation}
z(\rho,\theta,F)=A(F)({\alpha}_{20}Z_{20}(\rho,\theta)+{\alpha}_{22}Z_{22}(\rho,\theta)+{\alpha}_{31}Z_{31}(\rho,\theta)),
\label{eq:oap_comsa}
\end{equation}
with $A(F)$ a coefficient linking the amplitude of the deformation to the force $F$.\\
The system design is finally optimized with Finite Element Analysis, as described in Section \ref{sec:method_design}.
The Finite Element model performance is promising: with only one actuator, the required mirror shape is generated with high precision: 
the residual deformation is 14.4 nm rms for an optical surface shape of 2.7 $\mu$m rms (see Fig. \ref{fig:perfo_OAP}),
and it can generate the set of OAPs shown in Fig. \ref{fig:OAP_comsa} with this precision of 0.5\%.

\section{Conclusion}
The two concepts presented in this paper allow the generation of an optical mode with one actuator.
Moreover, the number of actuators does not depend on the mirror size and its action does not induce high spatial frequency errors.
The optical surface is deformed through the application of bending moments at the mirror edges.
The system geometry is relatively simple: the bending moments are generated by the application of forces on an
intermediate plate and they are transmitted to the mirror with a ring linking the plates together.
This warping structure is clamped on a reference plate through another ring.
The VOALA system is able to generate independently focus, astigmatism and coma with one actuator per mode.
The COMSA system has a single actuator, generating a given combination of optical aberrations.
Such optimized systems are designed to fill-in an optical function defined by a specific instrument design and
they will allow an efficient wave-front error compensation.
The expected performance of these active optics systems have been computed with Finite Element Analysis, 
showing excellent performance for visible/IR applications.
The next step is the manufacturing of prototypes to experimentally validate the simulation results.
We have already done this work on different deformable systems \cite{2009ApOpt..48.2932H,2012SPIE.madras}, 
allowing the validation of our Finite Element Analysis method and the correlation of simulated and experimental results.\\
With a minimum number of degrees of freedom, this type of system is convenient to use in many applications.
The next generation of Earth- and Space-based large lightweight segmented telescopes will benefit of simple
active mirrors, allowing an in-situ active shape error compensation with a minimal volume, weight and power consumption.
For instance, Patterson \cite{2010SPIE.7731E..60P} proposes a diluted, reconfigurable pupil telescope with a primary mirror 
constituted of several movable off-axis parabolas, whose asphericity depends on the required configuration.
Still in the field of large telescopes, these simple deformation devices, generating OAP, can be coupled to stress polishing technique
for the manufacturing of primary mirrors' segments.
Another interesting application for the VOALA and COMSA concepts lies in zoom systems, composed of two active mirrors. 
Wick has developed a zoom with two 59-channels active mirrors \cite{2006amos.confE..13W}.
In such a system, the mirrors need to generate focus for on axis beams and a combination of focus, astigmatism and coma for off-axis beams.
These functions would be perfectly fulfilled by our systems with no more than 3 actuators per mirror.
The compensation of field effects, non common path aberrations or optics deformation are also major applications:
a simple correcting deformable mirror in the optical train will allow significant constraints relaxation on systems 
assembly and integration as well as on operational stability.

\section*{Acknowledgments}
This work is performed with the support of a Ph.D grant from CNES (Centre National d'Etudes Spatiales) and Thales Alenia Space.

%%%%%%%%%%%%%%%%%%%%%%% References %%%%%%%%%%%%%%%%%%%%%%%%%


\begin{thebibliography}{10}
\newcommand{\enquote}[1]{``#1''}

\bibitem{2000eaa..bookE4780.}
P.~{Murdin}, \emph{{Active Optics}}, Encyclopedia of Astronomy and
  Astrophysics (2000).

\bibitem{1987JMO....34..485W}
R.~N. {Wilson}, F.~{Franza}, and L.~{Noethe}, \enquote{{Active optics. I. A
  system for optimizing the optical quality and reducing the costs of large
  telescopes.}}, Journal of Modern Optics., vol. 34, pp. 485--509 (1987).

\bibitem{1994SPIE.2199..271K}
E.-D. {Knohl}, \enquote{{VLT primary support system}}, Society of Photo-Optical Instrumentation Engineers
  (SPIE) Conference Series, vol. 2199, pp. 271--283 (1994).

\bibitem{1998A&AS..128..221F}
M.~{Ferrari}, \enquote{{Development of a variable curvature mirror for the
  delay lines of the VLT interferometer}}, Astronomy and Astrophysics,
  vol. 128, pp. 221--227 (1998).

\bibitem{Schmidt1932}
B.~{Schmidt}, \enquote{{A coma-free telescope}}, Mitt.Hamburg Sternv,  pp. 7--15 (1932).

\bibitem{1972ApOpt..11.1630L}
G.~{Lemaitre}, \enquote{{New procedure for making Schmidt corrector plates.}},
  Applied Optics, vol. 11, pp. 1630--1636 (1972).

\bibitem{1980ApOpt..19.2341N}
J.~E. {Nelson}, G.~{Gabor}, L.~K. {Hunt}, J.~{Lubliner}, and T.~S. {Mast},
  \enquote{{Stressed mirror polishing. 2: Fabrication of an off-axis section of
  a paraboloid}}, Applied Optics, vol. 19, pp. 2341--2352 (1980).

\bibitem{1998aoat.book.....H}
J.~W. {Hardy}, \emph{{Adaptive Optics for Astronomical Telescopes}}, Oxford University Press (1998).

\bibitem{1976JOSA...66..207N}
R.~J. {Noll}, \enquote{{Zernike polynomials and atmospheric turbulence}},
  Journal of the Optical Society of America, vol. 66, pp. 207--211
  (1976).

\bibitem{2011OptEn..50f3003C}
L.~E. {Cohan} and D.~W. {Miller}, \enquote{{Integrated modeling for design of
  lightweight, active mirrors}}, Optical Engineering, vol. 50, 063003
  (2011).

\bibitem{1998SPIE.3356..758R}
D.~C. {Redding}, S.~A. {Basinger}, A.~E. {Lowman}, A.~{Kissil}, P.~Y. {Bely},
  R.~{Burg}, R.~G. {Lyon}, G.~E. {Mosier}, M.~{Femiano}, M.~E. {Wilson}, R.~G.
  {Schunk}, L.~{Craig}, D.~N. {Jacobson}, J.~{Rakoczy}, and J.~B. {Hadaway},
  \enquote{{Wavefront sensing and control for a Next-Generation Space
  Telescope}},  Society of Photo-Optical Instrumentation Engineers (SPIE) Conference
  Series, vol. 3356, pp. 758--772 (1998).

\bibitem{2010SPIE.7739E.105L}
M.~{Laslandes}, M.~{Ferrari}, E.~{Hugot}, and G.~{Lemaitre},
  \enquote{{In-flight aberrations corrections for large space telescopes using
  active optics}}, Society of Photo-Optical
  Instrumentation Engineers (SPIE) Conference Series, vol. 7739 (2010).

\bibitem{2011SPIE.8169E...2L}
M.~{Laslandes}, N.~{Rousselet}, M.~{Ferrari}, E.~{Hugot}, J.~{Floriot},
  S.~{Viv{\`e}s}, G.~{Lemaitre}, J.~F. {Carr{\'e}}, and M.~{Cayrel},
  \enquote{{Stress polishing of E-ELT segment at LAM: full-scale demonstrator
  status}}, Society of Photo-Optical Instrumentation
  Engineers (SPIE) Conference Series, vol. 8169 (2011).

\bibitem{Timoshenko1959}
S.~P. {Timoshenko} and S.~{Woinowsky-Krieger}, \emph{{Theory of Plates and
  Shells}}, Engineering Mechanics Series (McGRAW-Hill International Editions, 1959).

\bibitem{Lemaitre2009}
G.~R. {Lema{\^i}tre}, \emph{{Astronomical Optics and Elasticity Theory - Active
  Optics Methods}}, Astronomy and Astrophysics Library (Springer, 2009).

\bibitem{1998ApOpt..37.4663D}
J.~C. {Dainty}, A.~V. {Koryabin}, and A.~V. {Kudryashov}, \enquote{{Low-Order
  Adaptive Deformable Mirror}}, Applied Optics, vol. 37, pp. 4663--4668 (1998).

\bibitem{1982ApOpt..21..580F}
R.~H. {Freeman} and J.~E. {Pearson}, \enquote{{Deformable mirrors for all
  seasons and reasons}}, Applied Optics, vol. 21, pp. 580--588 (1982).

\bibitem{VOALA}
M.~{Laslandes}, E.~{Hugot}, M.~{Ferrari}, and A.~{Liotard}, \enquote{{Mirror
  with mechanical device to generate optical aberrations}}, Patent Pending
  (TAS, CNES, CNRS and Universit\'e de Provence), FR1102805 (2011).

\bibitem{COMSA}
M.~{Laslandes}, E.~{Hugot}, and M.~{Ferrari}, \enquote{{Correcting device with
  a deformable mirror for the compensation of at least one aberration with a
  known evolution}}, Patent Pending (CNES, TAS, CNRS and Universit\'e de
  Provence), FR1153390 (2011).

\bibitem{Smith2004}
I.~{Smith} and D.~{Griffiths}, \emph{{Programming the Finite Element Method}},
  4th edition, Wiley (2004).

\bibitem{Bonnans2009}
J.~{Bonnans}, J.~{Gilbert}, C.~{Lemarechal}, and C.~{Sagastizabal},
  \emph{{Numerical optimization: theoretical and practical aspects}}, Part I,
  Springer (2009).

\bibitem{1980ApOpt..19.2332L}
J.~{Lubliner} and J.~E. {Nelson}, \enquote{{Stressed mirror polishing. 1: A
  technique for producing nonaxisymmetric mirrors}}, Applied Optics, vol. 19, pp.
  2332 (1980).

\bibitem{2008ApOpt..47.1401H}
E.~{Hugot}, G.~R. {Lema{\^i}tre}, and M.~{Ferrari}, \enquote{{Active optics:
  single actuator principle and angular thickness distribution for astigmatism
  compensation by elasticity}}, Applied Optics, vol. 47, pp. 1401--1409 (2008).

\bibitem{2009ApOpt..48.2932H}
E.~{Hugot}, M.~{Ferrari}, K.~E. {Hadi}, P.~{Vola}, J.~L. {Gimenez}, G.~R.
  {Lemaitre}, P.~{Rabou}, K.~{Dohlen}, P.~{Puget}, J.~L. {Beuzit}, and
  N.~{Hubin}, \enquote{{Active Optics: stress polishing of toric mirrors for
  the VLT SPHERE adaptive optics system}}, Applied Optics, vol. 48, pp. 2932
  (2009).

\bibitem{2012SPIE.madras}
M.~{Laslandes}, C.~{Hourtoule}, E.~{Hugot}, M.~{Ferrari}, , C.~{Lopez},
  C.~{Devilliers}, A.~{Liotard}, and F.~{Chazallet}, \enquote{{Space active
  optics: performance of a deformable mirror for in-situ wave-front correction
  in space telescopes}}, Society of Photo-Optical
  Instrumentation Engineers (SPIE) Conference Series, vol. 8442 (2012).

\bibitem{2010SPIE.7731E..60P}
K.~{Patterson}, S.~{Pellegrino}, and J.~{Breckinridge}, \enquote{{Shape
  correction of thin mirrors in a recongurable modular space telescope}}, Society of Photo-Optical Instrumentation Engineers
  (SPIE) Conference Series, vol. 7731 (2010).

\bibitem{2006amos.confE..13W}
D.~{Wick}, B.~{Bagwell}, T.~{Martinez}, D.~{Payne}, S.~{Restaino}, and
  R.~{Romeo}, \enquote{{Lightweight, Active Optics for Space and Near Space}},
  in The Advanced Maui Optical and Space Surveillance Technologies
  Conference (2006).

\end{thebibliography}
\end{document}